\documentclass[aps,preprint]{revtex4}%
\usepackage{amsfonts}
\usepackage{amsmath}
\usepackage{amssymb}
\usepackage{graphicx}%
\providecommand{\U}[1]{\protect\rule{.1in}{.1in}}

\begin{document}
\title[ ]{Disguised Electromagnetic Connections in Classical Electron Theory}
\author{Timothy H. Boyer}
\affiliation{Department of Physics, City College of the City University of New York, New
York, New York 10031, USA}

\begin{abstract}
In the first quarter of the 20th century, physicists were not aware of the
existence of classical electromagnetic zero-point radiation nor of the
importance of special relativity. \ Inclusion of these aspects allows
classical electron theory to be extended beyond its 19th century successes.
\ Here we review spherical electromagnetic radiation modes in a
conducting-walled spherical cavity and connect these modes to classical
electromagnetic zero-point radiation and to electromagnetic scale invariance.
\ Then we turn to the scattering of radiation in classical electron theory
within a simple approximation. \ We emphasize that, in steady-state, the
interaction between matter and radiation is disguised so that the mechanical
motion appears to occur without the emission of radiation, even though the
particle motion is actually driven by classical electromagnetic radiation.
\ It is pointed out that, for \textit{nonrelativistic} particles, only the
harmonic oscillator potential taken in the low-velocity limit allows a
consistent equilibrium with classical electromagnetic zero-point radiation.
\ For \textit{relativistic} particles, only the Coulomb potential is
consistent with electrodynamics. The classical analysis places restrictions on
the value of $e^{2}/\hbar c$.

\end{abstract}
\maketitle

\section{Introduction}

\subsection{Classical Electron Theory}

Classical electron theory was introduced at the end of the 19th century in the
attempt to combine classical mechanics with classical electromagnetism. \ The
theory, developed prominently by H. A. Lorentz, considered point charges (with
the charge-to-mass ratio $e/\mathfrak{m}$ found by J. J. Thompson) in
mechanical potentials and also interacting with electromagnetic radiation.
\ The theory had a number of successes including providing explanations for
the Faraday effect, optical birefringence, and the normal Zeeman effect, none
of which involve Planck's constant $\hbar$. \ However, when combined with
other contemporary ideas of physics, the theory led to unacceptable ideas such
as the Rayleigh-Jeans law for thermal radiation equilibrium and to atomic
collapse for the nuclear model of the hydrogen atom. \ In the first quarter of
the 20th century, classical electron theory was gradually abandoned in favor
initially of old quantum theory and then modern quantum mechanics. \ The
transition to old quantum theory was described by Born in this way:
\textquotedblleft the stationary states of an atomic system shall be
calculated, as far as possible, in accordance with the laws of classical
mechanics, but the classical theory of radiation is
disregarded.\textquotedblright\cite{Born}

The \textquotedblleft disregard\textquotedblright\ of radiation has
occasionally caught the attention of classical-minded physicists. \ If one
wishes to continue using classical electron theory to describe atomic
phenomena, then the disregard of radiation represents an error. \ Indeed,
classical electron theory can be extended beyond the successes of the very
early 20th century by including two aspects which were not used by the
physicists of that period: 1) the presence of random classical electromagnetic
zero-point radiation, which involves Planck's constant $\hbar$, and 2) the
importance of special relativity. \ These two neglected aspects allow
classical theory to explain additional phenomena, including Casimir forces,
van der Waals forces, the decrease of specific heats at low temperatures,
diamagnetism, the Planck spectrum of thermal radiation, and the absence of
atomic collapse.\cite{B2019b}\cite{CandZ}

\subsection{Classical\ Electromagnetic Zero-Point Radiation}

Many physicist are unfamiliar with the idea of classical electromagnetic
zero-point radiation which provides a natural extension of traditional
classical electron theory.\cite{Understanding} \ Classical electron theory
consists of three basic ideas: 1) Maxwell's equations for the electromagnetic
fields with point charges as sources, 2) Newton's second law for the motion of
these charges due to forces, including forces due to electromagnetic fields,
and 3) boundary conditions on the differential equations appearing in parts 1)
and 2). \ Classical electromagnetic zero-point radiation appears in item 3) as
the source-free boundary condition on the electromagnetic fields in Maxwell's
equations; it corresponds to the electromagnetic fields which are already
present when the field sources start to act. \ The typical texts of classical
electromagnetism\cite{SandF}\cite{Griffiths}\cite{Jackson}\cite{Zangwill}
scarcely mention the source-free boundary conditions. \ Indeed, in our
undergraduate laboratory experiments on electricity and magnetism, there is no
mention of the radio waves and thermal radiation which are present in the room
while the students adjust the sources for their experiments. \ However, at the
level of atomic physics, the source-free fields may become important.
\ Classical electromagnetic zero-point radiation is random classical radiation
(like classical thermal radiation) with a Lorentz-invariant spectrum,
$U_{zp}\left(  \omega\right)  =const\times\omega$. \ This spectrum of energy
$U$ versus frequency $\omega$ is also scale invariant, and so does not pick
out a preferred length, or time, or energy; its presence is not immediately
obvious. \ Zero-point radiation is regarded as the zero-temperature limit of
blackbody radiation and can be treated with random phases just as was
classical thermal radiation. \ At positive temperature, blackbody radiation
has a different spectrum, and includes a special energy, length, and frequency
which are evident in the Wien displacement rule $T\lambda=const.$ \ In order
to match the experimentally observed Casimir forces due to random
electromagnetic fields at low temperature, the scale of classical
electromagnetic zero-point radiation is fixed so that $U_{zp}\left(
\omega\right)  =\left(  1/2\right)  \hbar\omega$ where $\hbar$ is Planck's
constant. \ This place is the one point where Planck's constant $\hbar$ enters
the extended version of classical electron theory.

\subsection{Disguised Electromagnetic Interactions}

It is a curious aspect of classical mechanical systems in classical
electromagnetic zero-point energy that, \textit{in equilibrium, the mechanical
system inherits the scale of motion involving }$\hbar$\textit{ from the
zero-point radiation, but the actual average motion is maintained as though
there were no interaction with radiation.} \ The interaction with zero-point
radiation is disguised by the average balance between radiation emission and
absorption. \ 

In the present article, we explore again the classical interaction between
charged particles in potentials and classical zero-point radiation. \ Here we
make use of spherical radiation modes, which will be unfamiliar to many
physicists whose radiation training includes only plane waves, but we deal
with much easier calculations involving not \textit{random} radiation but
rather \textit{coherent} radiation scattering, and we deal with only circular
particle orbits. \ We start with an adiabatically-invariant radiation spectrum
in a spherical cavity, and then consider the interaction of this radiation
with charged particles, especially charged particles in circular orbits.
\ Adiabatic invariance for a mechanical system refers to a quantity which
maintains its value under a slow change in one of the parameters of the
system. \ For example, the ratio of energy $U$ to frequency $\omega$ for a
harmonic oscillator is an adiabatic invariant $U/\omega=J$ under very slow
changes of the oscillator frequency. \ The assumption of an
adiabatically-invariant radiation spectrum mimics a crucial aspect of
zero-point radiation. \ The situation of \textit{coherent} radiation driving
at the particle's resonant frequency removes the \textit{fluctuation} aspects
which are present in the full interaction with zero-point radiation. \ What
remains is only the average values. \ However, even these average values are
of interest and are indeed striking. \ 

\subsection{Outline of the Article}

We start by discussing an adiabatically-invariant spectrum of classical
radiation in a spherical, conducting-walled cavity. \ If the phases between
the radiation modes are random, then such an adiabatically-invariant spectrum
provides an approximation to classical zero-point radiation, and, indeed,
becomes Lorentz-invariant zero-point radiation in the large-cavity limit.
\ The zero-point radiation spectrum contains the least possible information
since there is no basis for distinguishing any frequency; the spectrum has no
preferred reference frame, and no preferred length or time or energy.

Next we turn to the interaction between classical electromagnetic radiation
and a point charge in a potential. \ The familiar elementary case involves the
scattering of a plane wave by a small linear electric dipole oscillator
oriented along the $z$-axis. \ However, we simplify the situation to involve
driving at the \textit{resonant} frequency. Also, we point out that the plane
wave can be expanded as a sum over spherical multipole modes, and the dipole
scattering actually involves only the spherical electric multipole field of
order $l=1,m=0,$ the electric dipole field. \ All the rest of the plane wave
is unscattered.\ \ The energy of the oscillator is determined by the energy of
this one driving spherical multipole radiation mode. If we consider only this
one spherical multipole field and the oscillating particle, then, in
steady-state, the oscillator gives the false impression that it is a
frictionless mechanical oscillation which has no interaction with radiation;
the oscillatory motion continues but there is no average energy flow to or
from the oscillator and no average flow of radiation energy in any direction.

Next we consider a \textit{nonrelativistic} charged particle in a
\textit{circular} orbit in a classical central potential, driven (in dipole
approximation) by an electric spherical multipole field of order $l=1,m=1$.
\ Specification of the mechanical system requires information regarding the
mass $\mathfrak{m}$ and charge $e$ of the particle and the parameters $\kappa$
and $\mathfrak{n}$ of the central potential taken as $V\left(  r\right)
=\kappa r^{\mathfrak{n}}/\mathfrak{n}$ where $\mathfrak{n}$ is a numerical
constant. \ For charge $e$, the radiation field associated with a circular
orbit depends upon only the radius $r$ and frequency $\omega$ of the
mechanical orbit. \ Just as for the linear oscillator, the mechanical energy
of the particle in the circular orbit is determined by the radiation energy of
the driving electromagnetic field. \ We show that such a system is consistent
with an adiabatically-invariant electromagnetic radiation spectrum only for
the harmonic-oscillator potential, $V\left(  r\right)  =\kappa r^{2}/2,$ in
the case of an orbit of small velocity. \ 

In contrast with the \textit{nonrelativistic} particle situation, a
\textit{relativistic} charged particle of arbitrary mass $\mathfrak{m}$ in a
circular orbit of any non-zero radius in a \textit{Coulomb} potential is
indeed consistent with coherent electromagnetic radiation in an
adiabatically-invariant spectrum. \ Finally, it is pointed out that
steady-state motion for the classical charge puts limitations on the ratio
$e^{2}/\left(  \hbar c\right)  $ involving the charge $e$ of the particle and
the scale $\hbar$ of the driving zero-point radiation. \ Classical electron
theory starts at a more primitive level than quantum theory and so gives at
least some restrictions on the value of the fine structure constant.
\ Finally, we close with a brief discussion of the contrast in points of view
of quantum theory and of \ classical electron theory with classical
electromagnetic zero-point radiation.

\section{Multipole Radiation Modes in a Spherical Cavity}

\subsection{Electric Fields of Spherical Radiation Modes}

A spherical cavity of radius $R$ in a perfect conductor allows standing
radiation fields which are termed spherical electromagnetic field modes. \ For
a cavity of finite radius $R$, the radiation modes are discrete and can be
labeled by the discrete indices $n,l,m$. \ Although the spherical multipole
radiation modes for arbitrary values of $l$ and $m$ are treated in the older,
junior-level electromagnetism textbook of Slater and Frank,\cite{SandF} only
the dipole fields, $l=1$, appear in Griffiths' text,\cite{Griffiths} and most
students who encounter general spherical multipole fields do so in a
graduate-level text such as Jackson\cite{Jackson} or Zangwill.\cite{Zangwill}
\ We are using Gaussian units throughout our analysis.

There are both electric and magnetic spherical multipole modes. \ The electric
field inside the spherical cavity can be written as a sum over the multipole
fields,
\begin{equation}
\mathbf{E}(\mathbf{r},t)=\operatorname{Re}%
{\textstyle\sum\nolimits_{n=1}^{\infty}}
{\textstyle\sum\nolimits_{l=1}^{\infty}}
{\textstyle\sum\nolimits_{m=-l}^{m=l}}
\left(  \mathbf{E}_{nlm}^{\left(  E\right)  }(\mathbf{r},t)+\mathbf{E}%
_{nlm}^{\left(  M\right)  }(\mathbf{r},t)\right)  .
\end{equation}
The electric multipole fields $\mathbf{E}_{nlm}^{\left(  E\right)  }$\ of
(angular) frequency $\omega_{nl}^{E}=ck_{nl}^{E}$ are given by%
\begin{equation}
\mathbf{E}_{nlm}^{\left(  E\right)  }(\mathbf{r},t)=\frac{a_{nlm}^{E}%
}{-ik_{nl}^{E}}\exp\left[  -ick_{nl}^{E}t\right]  \nabla\times\left[
j_{l}\left(  k_{nl}^{E}r\right)  \mathbf{X}_{lm}\left(  \theta,\phi\right)
\right]  \label{Enlm}%
\end{equation}
and the magnetic multipole fields $\mathbf{E}_{nlm}^{\left(  M\right)  }$\ of
frequency $\omega_{nl}^{M}=ck_{nl}^{M}$ are given by
\begin{equation}
\mathbf{E}_{nlm}^{\left(  M\right)  }(\mathbf{r},t)=a_{nlm}^{M}\exp\left[
-ick_{nl}^{M}t\right]  \left[  j_{l}\left(  k_{nl}^{M}r\right)  \mathbf{X}%
_{lm}\left(  \theta,\phi\right)  \right]  ,
\end{equation}
where $a_{nlm}^{E}$ and $a_{nlm}^{M}$ are constants, $j_{l}$ is the spherical
Bessel function of order $l,$ and the vector spherical harmonic $X_{lm}$ is
given by
\begin{equation}
\mathbf{X}_{lm}(\theta,\phi)=\frac{1}{i\sqrt{l\left(  l+1\right)  }}%
\mathbf{r}\times\nabla Y_{lm}\left(  \theta,\phi\right)  . \label{Xlm}%
\end{equation}
There are magnetic fields associated with the multipoles which can be written
in analogous form,\cite{Jackson} but, for the radiation energy balance in our
subsequent analysis, these fields are not needed. \ 

\subsection{Frequencies of Standing Wave Modes in a Spherical Cavity}

The normal mode (angular) frequencies are given by $\omega=ck$ where wave
numbers $k_{nl}^{M}$ and $k_{nl}^{E}$ are related to the zeros of the Bessel
functions or derivatives of the Riccati-Bessel functions
respectively\cite{SandF153}
\begin{equation}
j_{l}\left(  k_{nl}^{M}R\right)  =0\text{ \ and \ }\left(  \frac{d}{dr}\left[
rj_{l}(k_{nl}^{E}r)\right]  \right)  _{r=R}=0. \label{zeros}%
\end{equation}
If the radius $R$ of the conducting-walled cavity is sufficiently large
compared to the frequency of interest so that $k_{nl}R>>l,$ then the spherical
Bessel functions can be approximated by their large-argument asymptotic forms
\begin{equation}
j_{l}\left(  x\right)  \approx\frac{1}{x}\sin\left(  x-\frac{l\pi}{2}\right)
, \label{large}%
\end{equation}
so that the zeros of the Bessel functions and of the Riccati-Bessel function
can be approximated by
\begin{equation}
k_{nl}^{M}\approx\left(  n+\frac{l}{2}\right)  \frac{\pi}{R}\text{ \ and
\ }k_{nl}^{E}\approx\left(  n+\frac{l+1}{2}\right)  \frac{\pi}{R}.
\end{equation}

\subsection{Energy in Spherical Radiation Fields}

The energy in a spherical mode is connected to the expansion coefficients
$a_{nlm}^{E}$ and $a_{nlm}^{M}$. \ The energy in a normal mode is equally
divided (on time average) between the electric and magnetic fields. \ Thus
taking the electric mode labeled by $nlm,$ the energy for large $R,$
$1<<k_{lm}^{E}R,$ is
\begin{align}
U_{nlm}^{(E)}  &  =%
{\textstyle\int}
d^{3}r\frac{1}{8\pi}(E^{2}+B^{2})=\frac{1}{4\pi}%
{\textstyle\int\nolimits_{0}^{R}}
drr^{2}%
{\textstyle\int}
d\Omega\frac{1}{2}\mathbf{B\cdot B}^{\ast}\nonumber\\
&  =\frac{\left\vert a_{nlm}^{E}\right\vert ^{2}}{8\pi}%
{\textstyle\int\nolimits_{0}^{R}}
drr^{2}\left[  j_{l}\left(  k_{nl}^{E}r\right)  \right]  ^{2}%
{\textstyle\int}
d\Omega\left\vert \mathbf{X}_{lm}\left(  \theta,\phi\right)  \right\vert
^{2}\approx\frac{\left\vert a_{nlm}^{E}\right\vert ^{2}R}{16\pi\left(
k_{nl}^{E}\right)  ^{2}},
\end{align}
where the large-argument asymptotic form in (\ref{large}) has been used for
the Bessel function and the standard normalization for the vector spherical
harmonic. \ 

For classical zero-point radiation, these expansion coefficients are chosen to
correspond to an average energy $U=\hbar\omega/2$ for each normal mode of
(angular) frequency $\omega$. \ Then we have%
\begin{equation}
U_{nlm}^{(E)}=\frac{\left\vert a_{nlm}^{E}\right\vert ^{2}R}{16\pi\left(
k_{nl}^{E}\right)  ^{2}}=\frac{1}{2}\hbar ck_{nl}^{E} \label{Urad}%
\end{equation}
or
\begin{equation}
\left\vert a_{nlm}^{E}\right\vert ^{2}=\frac{16\pi\left(  k_{nl}^{E}\right)
^{2}}{R}U_{nlm}^{(E)}=8\pi\left(  k_{nl}^{E}\right)  ^{3}\frac{\hbar c}{R}.
\label{aEnlm}%
\end{equation}
The energy in the magnetic multipole mode is analogous, $U_{nlm}^{(M)}=\hbar
ck_{nl}^{M}/2$. \ 

\subsection{Angular Momentum in Spherical Radiation Fields}

The electromagnetic fields in a spherical cavity carry not only energy but
also angular momentum. \ For large values of $R$, spherical multipole fields
have very specific ratios between energy and angular momentum in the
radiation; for a spherical multipole field of order $l,m$, and frequency
$\omega,$ the ratio between the $z$-component of angular momentum $L_{z}$ and
the energy $U$\ in the multipole field in a large cavity is always given
by\cite{Jackson750}
\begin{equation}
\frac{L_{z}}{U}=\frac{m}{\omega}. \label{LzUmw}%
\end{equation}
This restriction in the electromagnetic fields will then limit the
charged-particle systems which can be coupled consistently to electromagnetic
zero-point radiation. \ 

\subsection{Scaling in Classical Electromagnetism}

The spectrum of electromagnetic zero-point radiation has no preferred length,
or time, or energy, and so is invariant under an adiabatic compression or
under a change of scale which maintains the values of the speed of light in
vacuum $c$ and the charge of the electron $e$. \ This scaling aspect of
classical electromagnetism which maintains the values of $c$ and $e$ often
goes unrecognized, and represents a sharp break away from classical mechanics.
\ Thus, classical mechanics contains no fundamental constants, and so
mechanical systems can be chosen with arbitrary length dimensions, arbitrary
periods of oscillation, and with arbitrary mass. \ In contrast,
electromagnetic systems are sharply constrained. \ For example, a plane wave
in vacuum of wavelength $\lambda$ is immediately known to have a frequency
$c/\lambda$. \ This connection stands in contrast with that for a wave in a
mechanical medium where a sound wave of known wavelength can have various
velocities and frequencies depending upon the medium. \ Similarly, the
electrostatic potential energy between two electrons separated by distance $r$
is immediately known to be $e^{2}/r$ whereas mechanical systems can have
general potentials $V\left(  \mathbf{r}\right)  $ connecting energy and
separation. \ 

Classical electromagnetic zero-point radiation (like all of electromagnetism)
has been termed $\sigma_{ltU^{-1}}$-scale invariant since only one parameter
or scale can be freely varied while preserving the values of $c$ and
$e$.\cite{B1989b} \ Thus, if we start with a spherical radiation mode labeled
by $l$ and $m$ having frequency $\omega$ and energy $U$ in a spherical,
conducting-walled cavity of radius $R$, and change the cavity radius $R$
adiabatically by a factor of $\sigma$ so that $R\rightarrow R^{\prime}=\sigma
R\,$, then the spherical radiation mode has its frequency changed as
$\omega\rightarrow\omega^{\prime}=\omega/\sigma$, and its energy changed as
$U\rightarrow U^{\prime}=U/\sigma.$ \ This is what is meant by
\textquotedblleft the $\sigma_{ltU^{-1}}$-scaling of electromagnetic
systems;\textquotedblright\ the lengths, times, and energies are all connected
together under an adiabatic change or a change of scale. \ \ The situation of
adiabatic invariance is probably most familiar in connection with slowly
shortening the length of a mechanical simple pendulum by pulling the
supporting string through a hole.\cite{AtomicP} \ In the present case, we are
dealing with a spherical radiation mode which has harmonic-oscillator behavior
analogous to that of a small-amplitude pendulum.\ \ The \textit{spectrum} of
classical electromagnetic zero-point radiation (where $U=\hbar\omega/2$) is
invariant under $\sigma_{ltU^{-1}}$-scaling; thus, if all lengths, times, and
inverse energies are transformed by the same factor $\sigma\,$, then the
\textit{spectrum} remains the same despite individual modes being changed in
wave number, frequency, and energy. \ The individual modes are moved to new
roles in the spectrum, but the \textit{spectrum} is unchanged. \ The Coulomb
potential $V\left(  r\right)  =-e^{2}/r$ is also $\sigma_{ltU^{-1}}$-scale
invariant since the electronic charge $e$ is an unchanged fundamental constant
while the potential energy $V$ scales inversely as the length $r$,
$V\rightarrow V^{\prime}=V/\sigma$ when $r\rightarrow r^{\prime}=\sigma r$. \ \ 

\subsection{Forming an Adiabatically-Invariant Spectrum of Radiation}

All the radiation normal modes in a spherical conducting-walled cavity satisfy
the conditions given in Eq. (\ref{zeros}). \ We obtain an adiabatically
invariant spectrum if we require that the energy for each mode is proportional
to its frequency, $U=b\omega,$ where $b$ is a single constant, the same for
each mode in the cavity. \ In order to mimic zero-point radiation, we choose
$b=\hbar/2$, so that $U=b\omega=\hbar\omega/2$. \ Under an adiabatic change in
the cavity radius $R$, the ratio $U/\omega$ for each mode is preserved as a
constant, and therefore the adiabatic change does not alter the
proportionality constant in $U=b\omega=\hbar\omega/2$. \ The frequency
$\omega$ and the energy of each mode is changed, but the spectrum of radiation
in the cavity is invariant under adiabatic change in the radius $R$. \ In the
limit as the radius of the conducting cavity becomes ever larger,
$R\rightarrow\infty$, the density of radiation normal modes become ever closer
in frequency, and we arrive at a continuous, adiabatically-invariant spectrum
of radiation. \ The relative phases of the radiation modes at different
frequencies can be chosen randomly, and, in this fashion, we would obtain an
adiabatically-invariant spectrum of random radiation mimicking classical
electromagnetic zero-point radiation. \ In the limit $R\rightarrow\infty,$
this radiation spectrum becomes Lorentz-invariant classical zero-point
radiation. \ 

\section{Scattering by a Small Electric Dipole Oscillator}

\subsection{Familiar Oscillator Example}

We start out with a familiar elementary example which emphasizes that particle
oscillation can sometimes disguise the presence of the radiation which causes
the oscillation. \ A familiar elementary scattering calculation\cite{Garg} in
classical electrodynamics treats a small, one-dimensional electric dipole
oscillator $\mathbf{p}\left(  t\right)  =\widehat{z}ez\left(  t\right)  $,
involving a particle of mass $\mathfrak{m}$ and charge $e$\ in a harmonic
potential $V(z)=\kappa z^{2}/2$, oriented along the $z$-axis, and of natural
frequency $\omega_{0}=\sqrt{\kappa/m}$, which is driven in dipole
approximation by a plane wave of (angular) frequency $\omega$ traveling in the
$x$-direction, $\mathbf{E}\left(  x,y,z,t\right)  =\widehat{z}E_{0}\cos\left[
\left(  \omega/c\right)  x-\omega t\right]  .$ \ Newton's second law for the
motion of the particle then gives%
\begin{equation}
\mathfrak{m}\ddot{z}=-\mathfrak{m}\omega_{0}^{2}z+\mathfrak{m\tau}\dddot
{z}+eE_{0}\cos\left[  \omega t\right]  , \label{N2nd}%
\end{equation}
where the spring restoring force is $-\kappa z=-m\omega_{0}^{2}z,$ and the
radiation damping term involves the time $\tau=2e^{2}/(3\mathfrak{m}c^{3})$.
\ In steady-state motion, the electric dipole oscillates as
\begin{equation}
z\left(  t\right)  =\operatorname{Re}\frac{\left(  e/\mathfrak{m}\right)
E_{0}\exp\left[  -i\omega t\right]  }{-\omega^{2}+\omega_{0}^{2}-i\tau
\omega^{3}}=\operatorname{Re}\frac{\left(  e/\mathfrak{m}\right)  E_{0}%
\exp\left[  -i\omega t+i\delta_{10}^{E}\right]  }{\left[  \left(  -\omega
^{2}+\omega_{0}^{2}\right)  ^{2}+\left(  \tau\omega^{3}\right)  ^{2}\right]
^{1/2}}, \label{zoft}%
\end{equation}
where $\delta_{10}^{E}$ is the phase shift\cite{Jackson773} between the
incident and scattered waves with
\begin{equation}
\tan\left(  \delta_{10}^{E}\right)  =\frac{\tau\omega^{3}}{\left(  -\omega
^{2}+\omega_{0}^{2}\right)  }. \label{phase}%
\end{equation}
(Here the notation $\delta_{10}^{E}$ refers to the \textit{electric} multipole
mode of order $l=1,m=0$.)

\subsection{Driving the Oscillator at the Resonant Frequency}

If the oscillator is driven at resonance where $\omega=\omega_{0}$, then the
particle mass $\mathfrak{m}$ cancels between the factor of $e/\mathfrak{m}$ in
the numerator of Eq. (\ref{zoft}) and the time $\tau$ in the denominator, the
phase shift $\delta_{10}^{E}$ in Eq. (\ref{phase}) goes to $\pi/2$, and the
displacement becomes%
\begin{equation}
z\left(  t\right)  =\frac{3c^{3}E_{0}}{2e\omega_{0}^{3}}\sin\left[  \omega
_{0}t\right]  . \label{zoftres}%
\end{equation}
The oscillation at $\sin\left[  \omega_{0}t\right]  $ is completely out of
phase with the driving radiation $E_{0}\cos\left[  \omega t\right]  $. \ The
amplitude of the oscillation is given by $z_{0}=3c^{3}E_{0}/\left(
2e\omega_{0}^{3}\right)  $. \ Although the emission and absorption of power
are out of phase, the power emitted as radiation by the dipole oscillator is
balanced over one oscillation cycle by the power delivered to the oscillator
by the driving plane wave%
\begin{equation}
\left\langle \left[  2e^{2}/\left(  3c^{3}\right)  \right]  \omega_{0}%
^{4}z_{0}^{2}\sin^{2}\left[  \omega_{0}t\right]  \right\rangle =\left\langle
ez_{0}\omega_{0}E_{0}\cos^{2}\left[  \omega_{0}t\right]  \right\rangle
\end{equation}
since $\left\langle \sin^{2}\left[  \omega_{0}t\right]  \right\rangle
=\left\langle \cos^{2}\left[  \omega_{0}t\right]  \right\rangle =1/2$. \ 

\subsection{Scattering Involves the Electric Spherical Radiation Mode
$l=1,m=0$}

The plane wave $\mathbf{E}\left(  x,y,z,t\right)  =\widehat{z}E_{0}\cos\left[
\left(  \omega/c\right)  x-\omega t\right]  $ can be expanded in terms of
spherical multipole modes.\cite{Jackson769} \ Although the traditional dipole
scattering problem is phrased in terms of plane waves (which are familiar at
the level of introductory physics and also appropriate for laboratory
experiments), the interaction between the small electric dipole and radiation
actually involves \textit{only} the electric spherical multipole radiation
mode $l=1,m=0$. \ Only the electric \textit{dipole} multipole fields,
$l=1,m=-1,0,1,$ are finite in the limit as the displacement $r$ becomes very
small, $r\rightarrow0$, because, for small argument, the spherical Bessel
functions go as the $l$th power of the argument,%

\begin{equation}
j_{l}\left(  x\right)  \approx\frac{x^{l}}{\left(  2l+1\right)  !!}.
\end{equation}
For a small electric dipole oscillating along the $z$-axis, only the
$\mathbf{E}_{lm}^{E}$ multipole given in Eq. (\ref{Enlm}) for $l=1,m=0$ will
drive the oscillator. \ Thus, at the natural frequency of the dipole
oscillator $\omega_{0}=ck_{n1}^{E}$ and for $\mathbf{X}_{10}\left(
\theta,\phi\right)  =\widehat{\phi}i\sqrt{3/\left(  8\pi\right)  }\sin\theta$,
this field becomes%
\begin{align}
\mathbf{E}_{10}^{E}(0,t)  &  =\operatorname{Re}\frac{a_{n10}^{E}}{-ik_{n1}%
^{E}}\exp\left[  -ick_{n10}^{E}t\right]  \nabla\times\left[  \frac{k_{n1}%
^{E}r}{3}\widehat{\phi}i\sqrt{\frac{3}{8\pi}}\sin\theta\right] \nonumber\\
&  =-\widehat{z}\frac{a_{n10}^{E}}{\sqrt{6\pi}}\cos\omega_{0}t
\end{align}
and is non-vanishing at the center of the dipole oscillation. \ This field
agrees with driving by the plane wave when we take
\begin{equation}
E_{0}=-\frac{a_{n10}^{E}}{\sqrt{6\pi}}=-\frac{1}{\sqrt{6\pi}}\sqrt{\frac
{16\pi\left(  k_{nl}^{E}\right)  ^{2}}{R}U_{nlm}^{(E)}} \label{E0U}%
\end{equation}
and the oscillator motion is given in Eq. (\ref{zoftres}). \ The oscillation
amplitude of the small electric dipole is%
\begin{equation}
p_{0}=ez_{0}=\frac{3c^{3}}{2\omega_{0}^{3}}E_{0}=-\frac{3c^{3}}{2\omega
_{0}^{3}}\frac{a_{n10}^{E}}{\sqrt{6\pi}}. \label{p0a}%
\end{equation}

\subsection{Interaction with a Single Radiation Mode}

We can imagine that the small electric dipole oscillator is located at the
center of a spherical cavity with conducting walls containing an
adiabatically-invariant spectrum of radiation, corresponding to that discussed
earlier which forms a parallel with classical zero-point radiation. \ For a
cavity of finite radius $R$, the mass $\mathfrak{m}$ of the particle can be
chosen so large that the line width $\Gamma=2e^{2}\omega^{2}/\left(
3mc^{3}\right)  $ is smaller than the separation in frequency between the
discrete radiation modes of the cavity, and the particle interacts at its
resonant frequency with a single mode of radiation which can be regarded as
coherent radiation. \ As the radius $R$ increases toward infinity, the number
of normal modes per unit frequency interval increases in compensation. \ In
order to interact with a single frequency, the ratio $e/\mathfrak{m},$
involving the charge $e$ divided by the particle mass $\mathfrak{m}$, must
become ever smaller as the radius $R$ of the cavity increases and the
separation between the modes decreases. \ 

\subsection{Analogue with Radiationless Stationary States}

If we consider the oscillator along the $z$-axis as driven, not by a plane
wave, but by only the part of the plane wave with which it actually interacts,
then we have a situation involving an oscillator in a standing spherical wave
field $\mathbf{E}_{10}^{E}(\mathbf{r},t)$. \ The electromagnetic wave forces
the dipole into oscillation, and the oscillator then radiates into exactly the
same spherical multipole mode, but with a phase shift of $\pi/2$. \ In the
steady-state situation, the energy provided to the oscillator exactly balances
the energy emitted by the oscillator and gives a standing-wave radiation
pattern. \ In this situation, no experiment would be able to detect an average
flow of energy. \ There is an oscillating electromagnetic radiation field and
also a charged particle oscillating in a harmonic potential. \ However, there
is no average transfer of energy between the radiation and the charge. \ The
oscillation amplitude of the small electric dipole is determined by the
amplitude of the spherical multipole radiation field which maintains the
amplitude of the oscillation against radiation emission by the oscillator.
\ However, this driving is completely hidden by the radiation balance. \ This
situation is a model for the sort of interaction with classical zero-point
radiation which has been suggested\cite{Marshall}\cite{B1975} as possibly
providing a classical basis for the ground state of atomic hydrogen or of any
quantum harmonic oscillator system. \ 

\subsection{Consistency Under Adiabatic Change of Oscillator Frequency}

We have noted that zero-point radiation is adiabatically invariant. \ Here, it
is of interest to note that the connection between the oscillator motion and
the driving radiation is maintained during an adiabatic change of the
oscillator frequency when the oscillator is in an adiabatically-invariant
radiation spectrum. \ During an adiabatic change in the oscillator frequency,
the ratio of the oscillator energy $\mathcal{E}$ to the oscillation frequency
$\omega_{0}$ remains unchanged, $\mathcal{E}/\omega_{0}=const.$ \ But from
Eqs. (\ref{p0a}) and (\ref{Urad}), the oscillator energy is related to
the\ amplitude of the particle oscillation which is related to the radiation
driving force, and so to the radiation energy per normal mode%
\begin{equation}
\mathcal{E}=\frac{1}{2}\mathfrak{m}\omega_{0}^{2}z_{0}^{2}=\frac{1}{2}%
\frac{\mathfrak{m}}{e^{2}}\omega_{0}^{2}\left(  \frac{3c^{3}}{2\omega_{0}^{3}%
}\frac{a_{nlm}^{E}}{\sqrt{6\pi}}\right)  ^{2}=\left[  3\left(  \frac
{\mathfrak{m}c^{2}}{e^{2}}\right)  \frac{c^{2}}{\omega_{0}^{2}}\frac{1}%
{R}\right]  U.
\end{equation}
We will require that the factor of proportionality connecting the mechanical
energy $\mathcal{E}$ to the radiation mode energy $U$ is always one,
\begin{equation}
\left[  3\left(  \frac{\mathfrak{m}c^{2}}{e^{2}}\right)  \frac{c^{2}}%
{\omega_{0}^{2}}\frac{1}{R}\right]  =1\text{ \ or \ }\frac{1}{3}\left(
\frac{e^{2}}{\mathfrak{m}c^{2}}\right)  \frac{\omega^{2}}{c^{2}}R=1,
\label{condit}%
\end{equation}
so that the oscillator energy $\mathcal{E}$ equals the radiation mode energy
$U$. \ 

The situation here for single-mode driving is analogous to the
narrow-line-width approximation $\tau\omega_{0}<<1$ involving random driving
radiation. The quantity $\tau=2e^{2}/\left(  3\mathfrak{m}c^{3}\right)  $
gives a characteristic time associated with the charged particle, and the
quantity $\Gamma=\left(  2/3\right)  \left[  e^{2}/\left(  mc^{3}\right)
\right]  \omega_{0}^{2}$ gives a characteristic frequency spread. \ In this
article, we consider a connection between the mechanical oscillator and a
single radiation mode at the oscillator's resonant frequency. \ In earlier
work with random radiation, we integrated over the sharply peaked function
given in Eq. (\ref{zoft}) to obtain the average energy of the oscillator
$\left\langle \mathcal{E}\right\rangle $ as connected to the average energy of
the radiation $\left\langle U\left(  \omega_{0}\right)  \right\rangle $ at the
oscillator's resonant frequency $\omega_{0}$.\cite{Marshall}\cite{B1975} \ For
large cavity radius $R$, the radiation modes are closely spaced so that
$\Gamma$ would cover a large number of normal mode frequencies. \ Here we
require that $e^{2}/\left(  \mathfrak{m}c^{2}\right)  $ goes to zero and $R$
goes to infinity in such a way that the condition (\ref{condit}) always holds.
\ Then our one-mode-coherent-field calculations come close to the random
zero-point calculations; the energy for the oscillator matches the energy per
normal mode for the radiation modes at the resonant frequency of the
oscillator. \ \ In both our resonant treatment and the traditional
narrow-line-width calculation, the charge $e$ has entirely disappeared from
the problem. \ There appears to be no connection between the mechanical system
and the radiation system. \ Yet we have equality between the energy $U$ in the
radiation field and the energy\ $\mathcal{E}$\ of the mechanical system. \ The
situation looks like a mechanics problem in the fashion suggested by Born's
quote in the introduction where the radiation is to be "disregarded," except
that here we know the amplitude of the mechanical oscillation is actually
determined by the driving electromagnetic radiation. \ 

\section{Circular Particle Orbits in a Central Potential}

\subsection{Circular Orbits with Both Energy and Angular Momentum}

We now generalize the mechanical system under consideration. \ For the small
one-dimensional linear oscillator considered above, the oscillator energy is
the one quantity which determines the amplitude of the oscillation, and, for a
small electric dipole oscillator, this energy is determined by the amplitude
of the driving coherent radiation at resonance. \ We now turn to charged
particles in circular orbits in central potentials when driven by coherent
radiation. \ Although our ultimate, long-term interest is the motion of
charged particles in random classical zero-point radiation, the
coherent-radiation situation can already remind us of important aspects of the
connection between radiation and matter. \ A charged particle in a circular
orbit in a central potential has both energy and angular momentum, which it
must be radiating away. \ If the circular orbit is to be steady-state, then
there must be electromagnetic driving fields which supply energy and angular
momentum to the charge so as to balance the emitted energy and angular
momentum. \ 

\subsection{Circular Orbits for Nonrelativistic Particles}

Traditional classical electron theory treats nonrelativistic charged particles
in arbitrary potential functions $V\left(  \mathbf{r}\right)  $. \ Therefore,
we turn to more general potential functions, but restrict our analysis to
\textit{circular} orbits in central potentials $V\left(  r\right)  $. We
consider first charged particles of mass $\mathfrak{m}$ and charge $e$ in
\textit{nonrelativistic} circular orbits within a central potential $V\left(
r\right)  =\kappa r^{\mathfrak{n}}/\mathfrak{n}$. \ The particle motion lies
in a plane which can be taken as the $xy$-plane, $\mathbf{r}\left(  t\right)
=r\left[  \widehat{i}\cos\left(  \omega t\right)  +\widehat{j}\sin\left(
\omega t\right)  \right]  $. \ For the \textit{mechanical} motion of
nonrelativistic particles in such potentials, the charge $e$ is regarded as
irrelevant, and Newton's second law gives the balance for centripetal
acceleration as
\begin{equation}
\mathfrak{m}\frac{v^{2}}{r}=\kappa r^{n-1}\text{ \ \ or \ }\left(  \omega
r\right)  ^{2}=v^{2}=\frac{\kappa}{\mathfrak{m}}r^{n}.\text{ \ } \label{N2b}%
\end{equation}
Thus, in general, the ratio $\kappa/\mathfrak{m}$ provides the connection
between the particle speed $v$ and the orbital radius $r$. \ The energy
$\mathcal{E}$ of the particle in the circular orbit in the potential is given
by%
\begin{equation}
\mathcal{E}=\frac{1}{2}\mathfrak{m}v^{2}+\frac{\kappa r^{\mathfrak{n}}%
}{\mathfrak{n}}=\left(  \frac{n+2}{2n}\right)  \mathfrak{m}v^{2}. \label{Env}%
\end{equation}
The angular momentum $J_{\phi}$\ of the particle in its circular orbit is%
\begin{equation}
J_{\phi}=\mathfrak{m}r^{2}\omega=\frac{\mathfrak{m}v^{2}}{\omega}.
\label{Jphi}%
\end{equation}
Then taking the ratio of the particle angular moment to energy, we have
\begin{equation}
\frac{J_{\phi}}{\mathcal{E}}=\left(  \frac{2\mathfrak{n}}{\mathfrak{n}%
+2}\right)  \frac{1}{\omega}. \label{ratJE}%
\end{equation}
Indeed, since we have the general relation $\omega=\partial\mathcal{E}%
/\partial J_{\phi}$, the product $J_{\phi}\omega$\ of the orbital angular
momentum and the angular frequency has the same dimensions as the energy
$\mathcal{E}$. \ For our circular orbit in a central potential of the form
$V\left(  r\right)  =\kappa r^{\mathfrak{n}}/\mathfrak{n}$, the ratio
$J_{\phi}\omega/\mathcal{E}$ must be a dimensionless number. \ For our
nonrelativistic particle, we find from Eq. (\ref{ratJE}),
\begin{equation}
\frac{J_{\phi}\omega}{\mathcal{E}}=\left(  \frac{2\mathfrak{n}}{\mathfrak{n}%
+2}\right)  . \label{ratJoE}%
\end{equation}
Thus for a \textit{nonrelativistic} particle in a circular orbit, the ratio
$J_{\phi}\omega/\mathcal{E}$ is a dimensionless number which is characteristic
of the power $\mathfrak{n}$ of $r^{\mathfrak{n}}$ in the potential function
$V\left(  r\right)  =\kappa r^{\mathfrak{n}}/\mathfrak{n}$. \ 

\section{Connection of Circular Particle Orbits to Electromagnetic Radiation}

\subsection{Requirements on the Driving Radiation}

Since the particle in the potential $V\left(  r\right)  =\kappa
r^{\mathfrak{n}}/\mathfrak{n}$ is assumed charged, we are interested in
connecting our mechanical system with electromagnetic radiation.
\ Electromagnetic radiation is a wave of speed $c$, so that we expect the
\textit{nonrelativistic} particle speed $\omega r=v$ to be much less than the
radiation speed $c$, $v<<c$. \ Accordingly, the dipole approximation is
appropriate for the connection between the nonrelativistic mechanical system
and the radiation. \ The associated spherical radiation multipole coupled with
this electromagnetic current is the electric multipole mode of order
$l=1,m=1$, taken in the limit $\omega r/c=kr<<1$. \ 

In a cavity of radius $R$ for a particle of sufficiently large mass
$\mathfrak{m}$, the steady-state particle motion must be coupled to a single
radiation mode of frequency $\omega=\omega_{nlm}$ for $l=1,m=1$, with the
electric radiation field appropriately in phase with the particle orbital
velocity so as to provide exactly the energy which is radiated away by the
charged particle. \ Thus in this nonrelativistic dipole approximation, we have
the radiation balance condition that%

\begin{equation}
\frac{2e^{2}}{3c^{3}}\omega^{4}r^{2}=eE_{0}\omega r
\end{equation}
or%
\begin{equation}
E_{0}=\frac{2e}{3c^{3}}\omega^{2}v \label{E011v}%
\end{equation}
for $v<<c$. \ In this case, we need the electric field $E_{0}=E_{11\phi
}^{\left(  E\right)  }$ for small displacement $r$. \ Then using Eq.
(\ref{Xlm}) to obtain $X_{11}\left(  \theta,\phi\right)  =\sqrt{3/\left(
16\pi\right)  }\left(  \widehat{\theta}+\widehat{\phi}i\cos\theta\right)
\exp\left[  i\phi\right]  $, and substituting into Eq. (\ref{Enlm}) taken for
small values of $r$, we find $E_{11\phi}^{\left(  E\right)  }=a_{nlm}%
^{E}/\sqrt{12\pi}$. \ Then combining Eqs. (\ref{E011v}), (\ref{Env}), and
(\ref{E0U}), we have
\begin{align}
\mathcal{E}  &  \mathcal{=}\left(  \frac{\mathfrak{n}+2}{2\mathfrak{n}%
}\right)  \mathfrak{m}v^{2}=\left(  \frac{\mathfrak{n}+2}{2\mathfrak{n}%
}\right)  \mathfrak{m}\frac{3c^{4}}{e^{2}\omega^{2}}\frac{U}{R}\nonumber\\
&  =\left(  \frac{\mathfrak{n}+2}{2\mathfrak{n}}\right)  \left[  3\left(
\frac{\mathfrak{m}c^{2}}{e^{2}}\right)  \frac{c^{2}}{\omega^{2}}\frac{1}%
{R}\right]  U. \label{Eto2U}%
\end{align}
Thus the mechanical energy of the particle in a circular orbit in the
potential $V\left(  r\right)  =\kappa r^{\mathfrak{n}}/\mathfrak{n}$ is
directly related to the energy $U$ of the driving radiation mode at the
orbital frequency $\omega$. \ The connection between the mechanical and
electromagnetic energies involves the same factor in square brackets which
appeared for the linear oscillator in one dimension. \ If we apply the same
condition (\ref{condit}) required for the one-dimensional oscillator motion,
then here only the isotropic harmonic oscillator potential where
$\mathfrak{n}=2$ has a mechanical energy $\mathcal{E}$ which equals the
radiation energy $U$ in the driving mode. \ 

The harmonic-oscillator potential is also singled out by the ratio $J_{\phi
}\omega/\mathcal{E}$ in Eq. (\ref{ratJoE}). \ This ratio agrees with the
corresponding ratio $L_{z}\omega/U=1$ for the electromagnetic driving normal
mode of order $l=1,m=1$ \textit{only} when $\mathfrak{n}=2$, corresponding to
the harmonic oscillator potential $V\left(  r\right)  =\kappa r^{2}/2$,%
\begin{equation}
\frac{J_{\phi}\omega}{\mathcal{E}}=1\text{ \ (harmonic oscillator potential).}%
\end{equation}

For a \textit{nonrelativistic} particle in a Coulomb potential $V\left(
r\right)  =-e^{2}/r$ where $\mathfrak{n}=-1$, one finds%
\begin{equation}
\frac{J_{\phi}\omega}{\mathcal{E}}=-2\text{ \ \ (Coulomb potential). \ }
\label{ratioC}%
\end{equation}
Thus, for the \textit{nonrelativistic} charged particle in a Coulomb
potential, the ratio in Eq. (\ref{ratioC}) does \textit{not} agree with that
of the associated driving radiation where $L_{z}\omega/U=1$. \ 

\subsection{Radiation Equilibrium in Zero-Point Radiation}

We are interested not in the radiation equilibrium for a single mechanical
system in some specialized radiation arrangement, but rather in the
equilibrium of a general class of systems for varying mass $\mathfrak{m}$ in
adiabatically invariant radiation. \ We emphasize that for a given charge $e$,
the radiation equilibrium of an orbiting particle depends upon \textit{only}
the frequency $\omega$ and radius $r$ of the circular orbit or, equivalently,
the speed $v$ and frequency $\omega$ of the particle orbit. \ We now consider
the question of consistency under a $\sigma_{ltU^{-1}}$-scale change for both
the particle orbit and the associated driving radiation. \ The radiation
spectrum, which is assumed adiabatically-invariant, is unchanged under this
scale change. \ However, while the speed $v$ of the nonrelativistic particle
in its orbit and the angular momentum $J_{\phi}$ are both unchanged under this
scale change, the frequency $\omega$ of the orbit will be changed by a factor
of $1/\sigma$, $\omega\rightarrow\omega^{\prime}=\omega/\sigma$. \ From Eq.
(\ref{Jphi}), we see that
\begin{equation}
v^{2}=\frac{J_{\phi}\omega}{\mathfrak{m}}. \label{v2Jwm}%
\end{equation}
This means that in order to retain the connection to an
adiabatically-invariant radiation spectrum (where the length, time, and
inverse energy all change together), the change in mass $\mathfrak{m}$ and the
change in frequency $\omega$ must be connected. \ There are only two
potentials for which this situation actually holds. \ Using equations
(\ref{Env}) and (\ref{Jphi}), we see that the orbital frequency is connected
to the particle mass $\mathfrak{m}$ and orbital angular momentum $J_{\phi}$ as%
\begin{equation}
\omega=\frac{\partial\mathcal{E}}{\partial J_{\phi}}=\frac{\kappa
^{2/(\mathfrak{n}+2)}}{\mathfrak{m}^{\mathfrak{n}/\left(  \mathfrak{n}%
+2\right)  }}J_{\phi}^{\left(  \mathfrak{n}-2\right)  /\left(  \mathfrak{n}%
+2\right)  }. \label{omegaeJ}%
\end{equation}
Only for $\mathfrak{n}=-1,$ corresponding to the Coulomb potential $V\left(
r\right)  =-e^{2}/r$, does the factor $\mathfrak{m}$ appear in first power in
the numerator of Eq. (\ref{omegaeJ}). \ In this case, the constant $\kappa$
corresponds to the square of the electric charge, $\kappa=e^{2},$ and does not
change under a $\sigma_{ltU^{-1}}$-scale transformation, while
\begin{equation}
v=e^{2}/J_{\phi}\text{ \ \ (Coulomb potential),} \label{ve2J}%
\end{equation}
which does not involve the mass $\mathfrak{m}$ at all. \ The particle energy
is
\begin{equation}
\mathcal{E=-}\frac{me^{4}}{2J_{\phi}^{2}}\text{ \ \ (Coulomb potential).}
\label{Eme4J2}%
\end{equation}
In this case, both $\omega$ and $\mathfrak{m}$ indeed have the same
$\sigma_{ltU^{-1}}$-scaling behavior involving $1/\sigma$. \ \ 

Although \textit{only} the Coulomb potential allows a scaling of mass
$\mathfrak{m}$ in agreement with the scaling of frequency $\omega$, the
scaling situation can be avoided entirely by going to the large-mass limit in
Eq. (\ref{v2Jwm}). \ If the quantity $J_{\phi}\omega$ is fixed, the large-mass
limit $\mathfrak{m}\rightarrow\infty$ will give the small-velocity limit
$v\rightarrow0$, which is indeed required for nonrelativistic particle
behavior. \ But this is the situation for the harmonic oscillator potential
$V\left(  r\right)  =\kappa r^{2}/2$ where $\mathfrak{n}=2,~$and the
mechanical frequency is $\omega=\omega_{0}=\sqrt{\kappa/\mathfrak{m}}.$ \ In
this case, the spring constant $\kappa$ must increased along with
$\mathfrak{m}$ so as to hold the frequency $\omega_{0}$ fixed. \ We see in Eq.
(\ref{omegaeJ}) that the harmonic oscillator provides the only potential where
the frequency is entirely independent of $J_{\phi}.$ \ 

\subsection{Successful Harmonic Oscillator Calculations}

Most of the calculations connecting charged mechanical systems to radiation
involve linear oscillators in one spatial dimension. \ Indeed, Planck
considered such an oscillator in random classical radiation at the end of the
19th century and concluded that the oscillator came to equilibrium with the
radiation (in dipole approximation) when the energy of the oscillator matched
the energy per normal mode of the random radiation at the frequency of the
oscillator. \ It has been pointed out several times that adiabatic changes in
the natural frequency of a linear harmonic oscillator\cite{B2021b} or
adiabatic changes in the mechanical interaction frequencies between
oscillators\cite{Cole} will preserve the connection between the mechanical
frequencies and classical electromagnetic zero-point radiation in the dipole
approximation. \ All these one-dimensional calculations do not involve angular
momentum. \ However, the isotropic oscillator in three dimensions was
considered in connection with diamagnetism, and, indeed, the connection
between the mechanical system and radiation was found to be preserved in
classical electromagnetic zero-point radiation in the dipole
approximation.\cite{B2019d} \ 

The treatments of nonrelativistic charged particles in potentials other than
the harmonic oscillator potential involve systems which will not fit
consistently with classical electromagnetic zero-point radiation. \ Within
classical physics, we must go to relativistic systems if we hope to move
beyond the harmonic oscillator and its radiation connection at low velocity.
\ Most physicists seem unaware of this requirement. \ 

\section{Relativistic Charged Particle in the Coulomb Potential}

\subsection{Relativistic Mechanical Motion}

The no-interaction theorem of Currie, Jordan, and Sudarshan\cite{CJS} requires
that any \textit{relativistic} system involving an interaction between
classical charged particles other than through point collisions requires the
introduction of a field theory. \ The only familiar mechanical potential which
can be extended to a fully relativistic system is the Coulomb potential which
appears as part of classical electrodynamics. \ As noted above, only the
Coulomb potential has scaling properties which fit with the $\sigma_{ltU^{-1}%
}$-scaling which appears in electromagnetic systems.

Once again, we consider a charged particle in a circular orbit in a potential,
this time using relativistic analysis for the particle motion in the Coulomb
potential $V\left(  r\right)  =-e^{2}/r$. \ In this case, the particle charge
$e$ not only connects the particle to the radiation field, but also determines
the particle motion in the Coulomb potential. \ Newton's second law for
relativistic momentum gives for the force-balance of the centripetal
acceleration%
\begin{equation}
\mathfrak{m}\gamma\frac{v^{2}}{r}=\frac{e^{2}}{r^{2}}\text{ \ or \ }\left(
\frac{e^{2}}{\mathfrak{m}c^{2}}\right)  \frac{1}{r}=\gamma\beta^{2}
\label{Newtr}%
\end{equation}
where $\beta=v/c$ and $\gamma=\left[  1-\beta^{2}\right]  ^{-1/2}$. \ Here the
ratio $e^{2}/\left(  \mathfrak{m}c^{2}\right)  $ with the dimensions of length
provides the connection between the velocity ratio $\beta$ contained in
$\gamma\beta^{2}$ and the radius $r$. \ The energy of the particle in the
circular orbit in the Coulomb potential is%
\begin{equation}
U_{e}=\mathfrak{m}\gamma c^{2}-\frac{e^{2}}{r}=\mathfrak{m}\gamma
c^{2}-\mathfrak{m}\gamma v^{2}=\mathfrak{m}c^{2}\sqrt{1-\left(  \frac{v^{2}%
}{c^{2}}\right)  }=\mathfrak{m}c^{2}\sqrt{1-\beta^{2}}.
\end{equation}
The angular momentum of the particle in the circular orbit is%
\begin{equation}
J_{\phi}=r\mathfrak{m}\gamma v=\frac{e^{2}}{v}\text{ \ or \ }\frac{v}{c}%
=\beta=\frac{e^{2}}{J_{\phi}c}. \label{Jvc}%
\end{equation}
The relationship found here connecting the velocity $v$ to the angular
momentum $J_{\phi},$ $v=e^{2}/J_{\phi}$, holds for both relativistic and
nonrelativistic treatments of a particle in a circular orbit in the Coulomb
potential. \ The relativistic energy in the circular orbit can be rewritten as%
\begin{equation}
\frac{U_{e}}{\mathfrak{m}c^{2}}=\sqrt{1-\left(  \frac{e^{2}}{J_{\phi}%
c}\right)  ^{2}}=\frac{1}{\gamma}, \label{Umc2J}%
\end{equation}
giving the angular frequency
\begin{equation}
\omega=\frac{\partial U}{\partial J_{\phi}}=\frac{\mathfrak{m}c^{3}}{e^{2}%
}\left(  \frac{e^{2}}{J_{\phi}c}\right)  ^{3}\left[  1-\left(  \frac{e^{2}%
}{J_{\phi}c}\right)  ^{2}\right]  ^{-1/2},
\end{equation}
or%
\begin{equation}
\left(  \frac{e^{2}}{\mathfrak{m}c^{3}}\right)  \omega=\gamma\beta^{3}.
\label{emcomega}%
\end{equation}
Since $\omega r/c=\beta,\ $the equation (\ref{emcomega}) is consistent with
Eq. (\ref{Newtr}). \ Thus the particle mass $\mathfrak{m}$ (when combined with
the fundamental invariants $e$ and $c$) provides a scale for all the
dimensions (length, time, and energy) connected to the mechanical motion of a
relativistic particle in the Coulomb potential. \ Under a change in the
particle mass $\mathfrak{m}$, this mechanical system follows the
$\sigma_{ltU^{-1}}$-scaling which holds for classical electromagnetism. \ 

\subsection{Ratio of Angular Momentum to Energy}

Once again, we would like to calculate the ratio of angular momentum times
angular frequency to energy for this relativistic situation. \ We find%
\begin{equation}
\frac{J_{\phi}\omega}{U}=\left(  \frac{e^{2}}{J_{\phi}c}\right)  ^{2}\left[
\sqrt{1-\left(  \frac{e^{2}}{J_{\phi}c}\right)  ^{2}}\right]  ^{-2}%
=\frac{\beta^{2}}{1-\beta^{2}}. \label{JowU}%
\end{equation}
We notice that here for the \textit{relativistic} Coulomb case, the ratio
$J_{\phi}\omega/U$ is dimensionless, but, in sharp contrast with the
nonrelativistic Coulomb case in (\ref{ratioC}) where $J_{\phi}\omega/U=-2$, is
not a constant. \ Nonrelativistic mechanics contains no fundamental constants
and the dimensionless ratio in Eq. (\ref{ratJoE}) involves the dimensionless
index $\mathfrak{n}$ which appears in the potential function $V\left(
r\right)  =\kappa r^{\mathfrak{n}}/\mathfrak{n}$. \ 

For a circular orbit of a relativistic particle in a Coulomb potential
$V\left(  r\right)  =-e^{2}/r$, the ratio $J_{\phi}\omega/U$ in Eq.
(\ref{JowU}) can range between 0 and infinity, as the speed ratio $\beta=v/c$
ranges between 0 and 1. \ In the \textquotedblleft nonrelativistic
limit\textquotedblright\ of the relativistic expression, we would need the
energy $\mathcal{E}$ above the rest energy $\mathfrak{m}c^{2}$, $\mathcal{E}%
=U-\mathfrak{m}c^{2},$ giving%
\begin{align}
\frac{J_{\phi}\omega}{\mathcal{E}}  &  =\left(  \frac{J_{\phi}\omega}%
{U}\right)  \frac{U}{\left(  U-\mathfrak{m}c^{2}\right)  }=\left(  \frac
{\beta^{2}}{1-\beta^{2}}\right)  \frac{\left(  1-\beta^{2}\right)  ^{1/2}%
}{\left[  \left(  1-\beta^{2}\right)  ^{1/2}-1\right]  }\nonumber\\
&  =\beta^{2}\frac{1}{\left[  -\beta^{2}/2+...\right]  }\rightarrow-2\text{
\ for \ }\beta\rightarrow0.
\end{align}
Thus in the limit of small particle speed $v$\ compared to $c$, the expression
(\ref{JowU}) indeed goes over to the earlier result for a nonrelativistic
particle in a Coulomb potential as given in Eq. (\ref{ratioC}). \ 

For both a nonrelativistic particle and for a relativistic particle in a
Coulomb potential, we find $v=e^{2}/J_{\phi}$. \ Therefore the speed $v$ is
\textquotedblleft small\textquotedblright\ only for $J_{\phi}$
\textquotedblleft large.\textquotedblright\ \ However, nonrelativistic
mechanics contains no fundamental parameter involving speed, and all speeds
are possible, $0\leq v\,<\infty$. \ Similarly, there is no scale for angular
momentum, and all values of angular momentum are possible. \ Thus, in
nonrelativistic physics, there is no fundamental criterion for what is meant
by a \textquotedblleft small\textquotedblright\ or \textquotedblleft
large\textquotedblright\ speed. \ It is only when relativistic theory is
introduced, either in terms of relativistic mechanics or in terms of
relativistic radiation with the appearance of a fundamental speed $c,$ that a
criterion for a large or small speed appears. \ If the theory contains both
the speed $c$ and the electronic charge $e$, then there is also a fundamental
scale $e^{2}/c$ for angular momentum. \ 

\subsection{Connection of a Relativistic Particle to Radiation}

A relativistic particle in a circular orbit in the Coulomb potential allows an
entirely different connection to the radiation field from the situation for
nonrelativistic particles. \ We are no longer restricted to small particle
speeds in the isotropic harmonic oscillator potential coupled to radiation
through the dipole approximation. \ Now radiation modes for all values of $l$
and $m$ are allowed and indeed required. \ Burko\cite{Burko} has given an
analysis in terms of spherical multipole modes for the radiation emitted by a
particle in uniform circular motion at any speed $v<c$. \ We can use his
results, but the radiation pattern is now chosen as a standing wave pattern,
rather than merely the emission of radiation as considered by Burko. \ Indeed,
we may consider the radiation field as involving standing radiation modes in
infinite space. \ Since the magnitude of the electronic charge $e$ is fixed,
the dimensional units can now be scaled based upon the particle mass
$\mathfrak{m}$, giving an energy ratio $U/(\mathfrak{m}c^{2})$, a length ratio
$r/\left[  e^{2}/\left(  \mathfrak{m}c^{2}\right)  \right]  $, and a time
ratio $t/\left[  e^{2}/\left(  \mathfrak{m}c^{3}\right)  \right]  $. \ An
adiabatic change in the invariant (rest) mass $\mathfrak{m}$ of the particle
in the circular orbit will give a change in the energy $U$, the radius
$r$,\ and frequency $\omega$ of the orbit, leaving the velocity of the
particle unchanged. \ The change in the energy $U,$ radius $r$, and frequency
$\omega$ can all be compensated by an adiabatic change in the frequency of the
driving radiation. \ However, if the driving radiation is adiabatically
invariant, the spectrum is unchanged by any adiabatic change or any scale
change. \ Both the mechanical system and the radiation involve the same
one-parameter $\sigma_{ltU^{-1}}$-connection between the length, time, and
energy under adiabatic changes. \ Indeed, the harmonic oscillator potential in
the small-velocity limit and the relativistic Coulomb potential are the only
classical scattering systems which preserve the spectrum of classical
electromagnetic zero-point radiation.\cite{B2021b} \ 

\section{Relativistic Hydrogen and Limits Involving $e^{2}/c$ and $\hbar$}

\subsection{Nonrelativistic Suggestions for the Hydrogen Ground State}

It has long been suggested that classical electromagnetic zero-point radiation
might account for the ground state of hydrogen by providing radiation pick-up
which balances the acceleration-related radiative energy loss. \ In 1975, a
qualitative calculation\cite{B1975} was presented based upon the radiative
energy loss of a nonrelativistic charge in a circular orbit in a Coulomb
potential%
\begin{equation}
\frac{d\mathcal{E}_{loss}}{dt}=\frac{2e^{2}}{3c^{3}}\omega^{4}r^{2}%
=\frac{2e^{6}}{3\mathfrak{m}^{2}c^{3}r^{4}},
\end{equation}
and the energy pick-up associated with a dipole rotor in classical zero-point
radiation as
\begin{equation}
\frac{d\mathcal{E}_{gain}}{dt}=\frac{e^{2}\hbar\omega^{3}}{2\mathfrak{mc}^{3}%
}=\frac{e^{5}\hbar}{2\mathfrak{m}^{5/2}c^{3}r^{9/2}},
\end{equation}
leading to an orbital radius $r=\left(  3/4\right)  ^{2}\hbar^{2}/\left(
me^{2}\right)  $. \ \ Although it was recognized at that time that use of
pick-up by an electric dipole oscillator, in contrast to use of a dipole
rotor, would give the exact result for the Bohr radius, the qualitative
analysis clearly did not justify any suggestion of exact agreement. \ In 2021,
a phase space calculation using action-angle variables for purely circular
nonrelativistic Coulomb orbits in classical zero-point radiation
suggested\cite{B2021a} a phase space probability distribution $P(J_{\phi
})=const\times\exp\left[  -J_{\phi}/\left(  \hbar/2\right)  \right]  $.
\ However, such a phase space distribution will lead to a divergent result for
the average nonrelativistic energy of the charged particle. \ Neither of these
nonrelativistic calculations suggests any limit involving Planck's constant
$\hbar.$ \ 

In \textit{nonrelativistic} classical electromagnetic theory, Planck's
constant $\hbar$ appears as the scale of zero-point radiation and may be
chosen as large or as small as one desires. \ A nonrelativistic charged
particle in a Coulomb potential $V\left(  r\right)  =-e^{2}/r$ involves only
the mechanical parameters of charge $e$ and particle mass $\mathfrak{m}$.
\ The presence of classical electromagnetic zero-point radiation introduces
the constant $\hbar$ associated with the scale of the random radiation and
having the dimensions of angular momentum. \ Thus the nonrelativistic
classical theory of a charged particle in a Coulomb potential in zero-point
radiation involves the fundamental parameters $e,\mathfrak{m},\hbar\,\ $and so
allows exactly one length $\hbar^{2}/\left(  \mathfrak{m}e^{2}\right)  ,$ one
time $\hbar^{3}/\left(  \mathfrak{m}e^{4}\right)  $ , and one energy
$\mathfrak{m}e^{4}/\hbar^{2}$. \ There exists no restriction on the values of
$e$, $\mathfrak{m}$, or $\hbar$. \ \ In steady state, if such exists in
nonrelativistic classical theory, we expect no restrictions on the average
values of orbital radius, or frequency, or of energy. \ Rather, we should find
actual values proportional to the corresponding length, time, and energy
formed from $e$, $\mathfrak{m}$, and $\hbar$. \ And indeed the nonrelativistic
qualitative calculations are consistent with this expectation.

\subsection{Relativistic Mechanics and a Lower Limit on the Value of $J_{\phi
}$}

The situation involving a charge particle in a Coulomb potential is changed
significantly by the introduction of relativity for the particle behavior.
\ Now the \textit{mechanical} system has parameters involving $e,\mathfrak{m}%
,$ and $c.$ Accordingly, from the \textit{mechanical} parameters alone, we may
form a length $e^{2}/\left(  \mathfrak{m}c^{2}\right)  ,$ a time
$e^{2}/\left(  \mathfrak{m}c^{3}\right)  ,$ and an energy $\mathfrak{m}c^{2}$.
\ Furthermore, we see from Eq. (\ref{Umc2J}), that there is a limit on the
smallest allowed value for the angular momentum $J_{\phi}$ of a relativistic
particle in a Coulomb potential. \ Thus for our circular orbit, we must have
$e^{2}/J_{\phi}=v\,<c$, or $e^{2}/c<J_{\phi}$. \ Indeed, for a general orbit
(not necessarily circular) of a bound relativistic particle in a Coulomb
potential, we have\cite{Goldstein}%
\begin{equation}
\frac{U}{\mathfrak{m}c^{2}}=\left\{  1+\left[  \left(  J_{3}-J_{2}\right)
c/e^{2}+\sqrt{\left(  J_{2}c/e^{2}\right)  ^{2}-1}\right]  ^{-2}\right\}
^{-1/2},
\end{equation}
where the action variables are given by $J_{2}=J_{\phi}+J_{\theta}$, and
$J_{3}=J_{2}+J_{r}$, and again there is the lower limit $e^{2}/c$\ on the
value of $J_{2}\,\ $corresponding to the total angular momentum of the
particle. \ We see that as the angular momentum decreases to $e^{2}/c,$
$J_{\phi}\rightarrow e^{2}/c$ for the circular orbit, the particle speed in
Eq. (\ref{Jvc}) increases to $c$, and the particle total energy in
(\ref{Umc2J}) decreases to zero . \ Thus, the situation for the relativistic
particle is completely different from that of the nonrelativistic particle in
Eqs. (\ref{ve2J}) and (\ref{Eme4J2}) where, as the angular momentum decreases
to zero, $J_{\phi}\rightarrow0,$ the particle speed increases indefinitely,
$v\rightarrow\infty,$ while the particle energy decreases indefinitely,
$\mathcal{E\rightarrow-\infty}$. \ 

It is crucial to remember that the \textit{relativistic} particle trajectories
change dramatically when the value of the angular momentum gets near to
$e^{2}/c$.\cite{B2004} \ This aspect has nothing to do with Planck's constant
$\hbar$ which appears as the scale of the adiabatic spectrum of classical
zero-point radiation. \ Rather this aspect depends only upon the square-root
of the quantity $J_{2}^{2}-\left(  e^{2}/c\right)  ^{2}$associated with the
relativistic\ mechanical behavior of the particle in the Coulomb potential.
\ For steady-state particle behavior, values of angular momentum $J_{2}$
smaller than $e^{2}/c$ are forbidden.

\subsection{Suggestion of Relations Relating $e^{2}/c$ and $\hbar$}

Now for our \textit{circular} relativistic particle orbits described in Eqs.
(\ref{Newtr}), and (\ref{Jvc}), the driving radiation must be such as to
balance the relativistic radiation energy loss of the accelerating charge,%
\begin{equation}
P_{loss}=\frac{2e^{2}}{3c^{3}}\frac{\gamma^{4}v^{4}}{r^{2}}=\frac{2}{3}\left(
mc^{2}\right)  \left(  \frac{mc^{3}}{e^{2}}\right)  \left(  \frac{e^{2}%
}{J_{\phi}c}\right)  ^{8}\left[  1-\left(  \frac{e^{2}}{J_{\phi}c}\right)
^{2}\right]  ^{-3}, \label{Ploss}%
\end{equation}
where we require $e^{2}/c<J_{\phi}$. \ The energy gain, when treated as due to
coherent radiation, corresponds to power delivered by the radiation
$evE_{\phi}(r,\phi,0,t).$ \ However, in this case, we cannot assume that the
radius $r$ of the orbit is small. \ As an approximation, we may take only the
$l=1$, $m=1$ multipole field, giving%

\begin{equation}
P_{gain}\approx evE_{11\phi}^{\left(  E\right)  }(r,\phi,0,t)=eckr\sqrt
{\frac{3}{16\pi}}a_{11}^{E}\left\{  \frac{1}{kr}\frac{\partial}{\partial
r}\left[  rj_{1}\left(  kr\right)  \right]  \right\}  . \label{Pgain}%
\end{equation}
The crucial aspect remains that if the scale of the driving field given by
$a_{11}^{E}$ is too small, then the driving radiation giving Eq. (\ref{Pgain})
cannot balance the radiation emitted according to Eq. (\ref{Ploss}). \ Thus if
the scale $\hbar$ of the classical zero-point radiation were too small, say
less than $e^{2}/c$, then no stable hydrogen atom would be possible in
classical theory, since all the circular orbits would be in the forbidden
region. \ A relativistic classical electromagnetic analysis seems to suggest
limits on the value of the fine structure constant.

\section{Contrasting Outlooks in Classical and Quantum Theories}

\subsection{Separate Behavior or Coupled Interactions Between Radiation and
Matter?}

The classical point of view taken here stands in sharp contrast with the view
appearing in quantum theory. \ In the \textit{classical} view here, charged
particle motion can be in steady state, despite emitting electromagnetic
radiation, because of the presence of radiation which is driving the particle
motion. \ However, only very specific potentials (the harmonic oscillator for
a low-velocity particle and the Coulomb potential for a relativistic particle)
allow a consistent treatment in classical electromagnetic zero-point
radiation. In \textit{quantum} theory on the other hand, \textit{quantum
behavior is intrinsic for every system}, and a mechanical particle can exist
in its quantum ground state for \textit{any} arbitrary potential $V\left(
\mathbf{r}\right)  .$ \ The particle state can be described by a
time-independent wave function involving $\hbar$ even though the particle
experiences no interactions at all with electromagnetic radiation. \ \ 

The successes of 19th century classical electron theory, such as the Faraday
effect, optical birefringence, and the normal Zeeman effect, do not involve
the particle's interaction with any \textit{fundamental} radiation
distribution and do not involve Planck's constant $\hbar$. \ It is only with
the introduction of classical electromagnetic zero-point radiation, that
classical electron theory can be extended by detailed electromagnetic
calculations to explain some microscopic-level phenomena which do indeed
involve Planck's constant $\hbar$, such as Casimir forces, van der Waals
forces, the decrease of specific heats at low temperatures, diamagnetism, the
Planck spectrum of thermal radiation, and the absence of atomic collapse. \ \ 

As emphasized in the present article, the extended classical electron theory
sharply restricts the allowed potential functions for charged particles in
zero-point radiation. \ In general, the successful calculations of classical
electron theory with classical electromagnetic zero-point radiation involve
either free electromagnetic fields (such as for Casimir forces) or charged
particles in harmonic potentials which are driven by classical zero-point
radiation (such as for van der Waals forces). \ The classical calculations
involving free electromagnetic fields are exactly parallel to the quantum
counterparts; the average values for the energy per normal mode agree between
the classical and the quantum theory, and this agreement is sufficient to
account for the average Casimir forces.\cite{B1975} \ On the other hand, for
the interaction of dipole oscillator systems, the classical theory has the
more primitive starting point, involving charged point particles in radiation.
\ The classical theory leads to phase space distributions for the particles
which have direct connections with the quantum harmonic-oscillator probability
distributions.\cite{B2020a} \ The classical and quantum theories agree
regarding average values of forces but disagree regarding fluctuations.
\ Although the average values of the forces (where there is agreement) have
indeed been checked by experiment, the fluctuation aspects do not seem to have
been tested.

The \textit{quantum} point of view, which treats \textquotedblleft quantum
behavior\textquotedblright\ as something entirely different from classical
charged particles interacting with zero-point radiation, sometimes gives rise
to controversies. \ For example, some quantum physicists claim that Casimir
forces between materials are due to quantum zero-point radiation while others
insist that Casimir forces do not depend at all on quantum electromagnetic
zero-point radiation but rather are van der Waals forces due to interactions
arising from quantum charge fluctuations in the molecules of the
materials.\cite{contro} \ From the viewpoint of \textit{classical} electron
theory, such controversies make little sense since there must be a consistent
interaction between radiation and charged matter, and the same interaction
gives rise to both Casimir forces and to van der Waals forces.

\subsection{Nonrelativistic or Relativistic Hydrogen Atom?}

The contrasting descriptions for charged particles in the Coulomb potential
are also striking. \ For the Bohr model of old quantum theory, the
Schroedinger wave function of nonrelativistic quantum theory, and Dirac's
component wave function of relativistic quantum theory, the hydrogen ground
state in quantum theory is complete without any need for electromagnetic
radiation. This quantum viewpoint stands in sharp contrast with classical
electron theory with classical electromagnetic zero-point radiation where the
hydrogen ground state depends crucially upon the absorption of radiation
energy from the classical zero-point radiation; it is this energy absorption
which prevents atomic collapse. \ For \textit{nonrelativistic} classical
electron theory including classical zero-point radiation, the necessary
hydrogen calculations are incomplete but tantalizing, involving numerical
simulations\cite{CandZ} and qualitative analytic calculations.\cite{B1975}
\ As suggested in the present article, only the full calculation of a
relativistic particle in zero-point radiation is entirely consistent with
classical theory. \ And such a calculation remains to be done. \ The
\textit{relativistic} treatment of a classical mechanical particle in a
Coulomb potential includes novel aspects which remain unfamiliar to most
physicists.\cite{B2004} \ Such an analysis requires not the simplified model
of the present article involving coherent radiation acting on the particle at
resonance, but rather full random classical electromagnetic zero-point
radiation with a scale set by $\hbar$ acting on relativistic charged
particles. \ Indeed, even qualitative classical electromagnetic analysis
suggests restrictions on the value of $e^{2}/\left(  \hbar c\right)  .$ \ The
classical electromagnetic treatment stands in complete contrast with quantum
theory which assigns the same value $\hbar$ to every physical system, both
mechanical and electromagnetic, and then changes the rules of interaction
between radiation and matter so as avoid any restrictions on the value of the
fine structure constant. \ It remains to be seen just how much information
regarding the value of the fine structure constant will appear from a
completely classical calculation. \ \ 

\section{Acknowledgment}

The present article was stimulated by the work of Professor Daniel C. Cole
suggesting the possibility of classical excited states based upon coherent
classical radiation interacting with a classical nonrelativistic charged
particle in an elliptical orbit in a Coulomb potential. \ D. C. Cole,
\textquotedblleft Subharmonic resonance and critical eccentricity for the
classical hydrogen atomic system,\textquotedblright\ Eur. Phys. J. D
\textbf{72}, 200-214 (2018). \ I wish to thank Professor Michael C. Mackey,
Professor Brian R. La Cour, and Professor Daniel C. Cole for reading the
manuscript and making helpful suggestions. \ 

\subsubsection{}

\bigskip

\end{document}